\documentclass{article}

\usepackage{journal_abbr_mac}

\usepackage[T1]{fontenc}

\usepackage{graphicx}	
\usepackage{amsmath}	
\usepackage{amssymb}

\usepackage[utf8]{inputenc}
\usepackage[english]{babel}
 
\usepackage{natbib}
\setcitestyle{authoryear}
  
\begin{document}

\begin{center}

{\Large \bf Galactic Habitability Re-Examined: Indications of Bimodality}\\

N. Stojkovi\'c,$^{1,2}$,B. Vukoti\'c,$^{1}$, N. Martinovi\'c,$^{1}$, M.M. \'Cirkovi\'c$^{1,*}$ and M. Micic$^{1}$

{\small \emph{
$^{1}$Astronomical Observatory, Volgina 7, 11160 Belgrade, Serbia \\
$^{2}$Department of Astronomy, Faculty of Mathematics, University of Belgrade Studentski trg 16,
11000 Belgrade, Serbia\\
}
}

{\tt
}

\end{center}

\begin{abstract}
The problem of the extent of habitable zones in different kinds of galaxies is one of the outstanding challenges for contemporary astrobiology. In the present study, we investigate habitability in a large sample of simulated galaxies from the Illustris Project in order to at least roughly quantify the hospitality to life of different galactic types. The pioneering study of \citet{2015ApJ...810L...2D} is critically examined and some of its results are amended. In particular, we find a tentative evidence for a second mode of galactic habitability comprising metal-rich dwarfs similar to IC 225, LMC or M32. The role of the galactic environment and the observation selection effects is briefly discussed and prospects for further research on the topic outlined. 
\end{abstract}





\section{Introduction}

Habitability has quickly become the central concept of astrobiology \citep[e.g.,][]{2005ARA&A..43...31C,2005OLEB...35..555G,horneck2007complete,2012AREPS..40..597L,2016AsBio..16..561D}. While research interest is naturally concentrated upon the habitability of newly discovered Earthlike planets around nearby stars, there is also an emerging interest for habitability of other astrophysical systems, in particular galaxies as main building blocks of the universe. The fact that different spatial locations within a stellar system such as the Milky Way may be characterized by varying degree of habitability was already clear to Alfred Russel Wallace, co-discoverer of natural selection and a forerunner of astrobiology \citep{1903mpus.book.....W}. Since the onset of the astrobiological revolution in mid-1990s, our best understanding has been embodied in the concept of the Galactic Habitable Zone \citep[henceforth GHZ;][]{2001Icar..152..185G,2004Sci...303...59L}.  Early treatments of GHZ have been strongly influenced by the rare Earth hypothesis of \citet{2000rewc.book.....W}, which outlined a ``geocentric'' astrobiology based on the local properties of Earth and the Solar system, but also the Milky Way \citep[for a detailed criticism of some of the ``rare Earth'' claims, see][]{2012asla.book.....C}. Therefore, much of the subsequent theoretical work on habitability on large scales could be characterized as ``Milky-Way-centric'' \citep[e.g.,][]{2008OLEB...38..535C,2009IJAsB...8..121F,2010IJAsB...9...73F,2015AsBio..15..683M}. For an exception to this trend and somewhat prescient study which traces some of the ideas of the present work, see \citet{2012IJAsB..11..157S}. With the metallicity buildup being of crucial importance  to build exoplanets, some of the more recent studies put emphasis on chemical evolution \citep{2008SSRv..135..313P,2013RMxAA..49..253C,2014MNRAS.440.2588S,2017A&A...605A..38S}, with some of them investigating the habitability of M31 \citep{2013RMxAA..49..253C,2014MNRAS.440.2588S}. 

Since our astrophysical understanding of the structure, origin and evolution of galaxies has made dramatic strides in recent decades, especially with the advent of large-scale cosmological numerical simulations, it has been quite natural to apply those insights to the question of habitability. First applications of cosmological simulations in astrobiology have been the studies of \citet{2016MNRAS.459.3512V}  and \citet{2017IJAsB..16...60F}; for a review see also \citet{vukotic2017nbody}. These works showed how the Galactic Habitable Zone emerges as a consequence of large-scale trends of galactic evolution. Both primary studies have shown a certain shift from the original range of galactocentric distances in \citet{2001Icar..152..185G} toward more outward regions of the disc, and hence smaller mean metallicities. For instance, in the last snapshot of \citet{2016MNRAS.459.3512V}, corresponding to the most recent epoch, the maximal probability of finding a habitable planet is between $10$ and $15$ kpc from the galactic centre. In addition, the study of \citet{2017IJAsB..16...60F} shows that dwarf galaxies, such as the satellites of Milky Way and Andromeda, can have a significant density of habitable systems, to even greater degree than their larger gravitational hosts. This is indicative of a shift in opinion as to the central role of high metallicity in determining the degree of habitability. If the shift is real when we discuss stellar populations within a galaxy such as the Milky Way, it is only reasonable to generalize it to any consideration of a sample of galaxies as well: lower (average) metallicity galaxies will be more appreciated as potential habitats.

Clearly, generality of these ideas can only be checked by investigations, both theoretical and observational, of a large sample of galaxies. Two recent studies have attracted lots of attention in this respect. \citet{2015ApJ...810L...2D} use a model which links global properties of galaxies with the factors limiting their habitability. In particular, the destructive effects of cosmic explosions (supernovae, gamma-ray bursts, and perhaps magnetars, extreme stellar flares, and similar events) have been the focus of the attention. It is clear that these effects can cause extreme ecological damage to local planetary biospheres and consequently reduce habitability within large volumes of space. For a sampling of huge literature on the subject in the last more than 60 years, see \citet{krasovsky1957supernova}; \citet{Schindewolf1962neokatastrophismus}; \citet{1968Sci...159..421T}; \citet{1974Sci...184.1079R}; \citet{1976Natur.263..398W}; \citet{1978Natur.271..430H}; \citet{1995ApJ...444L..53T}; \citet{1997RaPC...49..299B,2001RaPC...60..297B}; \citet{1998PhRvL..80.5813D}; \citet{2002ApJ...566..723S}; \citet{2003ApJ...585.1169G}; \citet{2005ApJ...634..509T}; \citet{2006NJPh....8..120T}; \citet{2009IJAsB...8..183T}; \citet{2011Ap&SS.336..287B}; \citet{bougroura2011threshold}; \citet{2012MNRAS.423.1234S}; \citet{2014IJAsB..13..319M}. Most cosmic explosions are proportional to the star formation rate, which gives us a handle to compare irradiated volumes and consequent decrease of habitability in different galaxies. On that basis, \citet{2015ApJ...810L...2D} conclude that, in sharp contrast to \citet{2000rewc.book.....W} and \citet{2001Icar..152..185G}, giant ellipticals with very small present-day star formation rates and moderately low metallicities, like Maffei 1, are the best abodes of life in the universe in the present epoch.

Along similar lines, \citet{2018MNRAS.475.1829S} investigate the cosmic evolution of habitability with galaxy merger trees, the method naturally following the deployment of N-body simulations for astrobiological purposes by \citet{2016MNRAS.459.3512V}. \citeauthor{2018MNRAS.475.1829S} use the Millenium Simulation and its derivative ``millimil'' sample in 63 time intervals (or ``snaps'') starting from the earliest redshifts  to the present day. In addition to metallicity weighting, irradiations of potentially habitable planets by SNe, GRBs and the central AGN are considered in order to reach the volume-averaged habitability over prolonged intervals of time (like the fiducial value of 6 Gyr). \citeauthor{2018MNRAS.475.1829S} find a complex situation, with no simple answer to the question which galaxies are most habitable at present. Even galaxies of the same total mass and luminosity manifest clear differences regarding their possible habitability histories (changes in the habitable fraction of stellar mass over time). This study first indicates that the distribution of habitable fraction among stellar systems is  bimodal, being either smaller than $5\%$ or greater than about $30\%$ at present epoch. One way in which conclusions of \citeauthor{2018MNRAS.475.1829S} could be generalized consists in not relying upon the strong mass-metallicity relation imposed on objects in the Millenium Simulation, esp. in the low mass range (more on this below). In addition, the work of \citet{2018MNRAS.475.1829S} correctly emphasizes that strong metallicity dependence of the GRB rates presents an important, and possibly dominant detrimental factor from the point of view of habitability, especially for the most massive, highest metallicity systems.

Hence, we have essentially two major views on the habitability of galaxies so far: (i) the conventional view based to a large degree on the ``rare Earth'' thinking of Gonzalez, Ward, and Brownlee limiting life to large spiral discs analogous to the Milky Way Population I, and (ii) the radical view, emerging since about 2015, that it is mostly quiescent early-type galaxies and dwarfs more similar to the local Population II that are the best abodes for life. Details may vary from study to study, but this dilemma is quickly becoming one of the central issues of astrobiology. Therefore, further insights and analyses in this area are highly desirable. 

In the rest of this paper, we report on the analysis of a large sample of simulated galaxies from the flagship Illustris Project \citep{2014MNRAS.444.1518V} database with the intent of establishing evolutionary properties of galaxies relevant for habitability and therefore test the conjecture of \citet{2015ApJ...810L...2D} about the prevalence of early-type galaxies in the overall astrobiological landscape. This type of theoretical analysis of the large-scale cosmological simulation is both complementary to the previous study and necessary as a precursor to future observational surveys. In particular, we apply the general analytic formalism for habitability of \citet{2015ApJ...810L...2D} to simulated galaxies and their evolution and, in the process, identify a new category of astrobiological interesting objects. Particular emphasis will be put on one aspect of our analysis, namely the emergence of a subpopulation of galaxies with relatively small stellar masses and relatively high metallicities as alternative habitats to conventional large galaxies (both giant spirals like the Milky Way and giant ellipticals like Maffei 1, as suggested by \citeauthor{2015ApJ...810L...2D}). In contrast to the work of \citet{2018MNRAS.475.1829S}, we use a higher-resolution, more up-to-date cosmological simulation and focus on recent epochs (a few last snapshots) averaged over cosmologically relevant volume. While we are forced by the lack of appropriate observations farther afield to use examples of dwarf galaxies belonging to the Local Group below, we do not specifically follow Local Group analogues (as do Stanway et al.). 

The rest of the paper is organized as follows. After we describe methods and the sample in Section 2, high-habitability dwarfs are looked in more detail in Section 3. Concluding section gives a detailed discussion of several outstanding issues of the astrobiology of galaxies and points out some directions for further research.

\section{Simulation and samples}

In this paper we use dataset containing simulated subhaloes from the cosmological simulation Illustris, completed in late 2013. Illustris Project is a cosmological hydrodynamical simulation using moving-mesh code Arepo \citep{2012MNRAS.425.3024V}, which includes a comprehensive set of physical models needed for closely following formation and evolution of galaxies. Spatial volume included in simulation is $(106.5)^3$ Mpc$^3$. There are six runs of Illustris simulation: three with full baryonic physics models and three containing only cold dark matter. Each of the six runs contains 136 snapshots of space volume in different redshifts (from $z$ = 127 to $z$ = 0) out of which $134$ are accessible by the database, and for each snapshot there is an accompanying halo and subhalo catalogue, as well as the SubLink merger tree (or two merger trees if baryonic physics is neglected). We use subhaloes--which are associated with galaxies--from Illustris-1 because it has the best mass resolution of the runs which includes baryonic physics models. The subhalo catalogue we consider is one associated with redshift $z$ = 0, i.e., the current epoch. This catalogue contains $4366546$ subhaloes with minimal total mass (which includes dark matter, stellar, and gas phase mass) of the order $10^7$ solar masses, and maximal total mass of the order $1014$ solar masses. When we exclude subhaloes with all three zero values for stellar mass, star formation rate (SFR), and metallicity (more precisely, its non-logaritmic form), we have $335722$ subhaloes ($7.7\%$ of the catalogue) in which formation of potentially habitable planets could have occurred. We consider only subhaloes with stellar mass greater than $10^7$ solar masses, which leaves our dataset with $148414$ subhaloes ($3.4\%$ of the catalogue). Mathematical considerations associated with calculating number of potentially habitable planets per procedure of \citet{2015ApJ...810L...2D} demand that we do not consider subhaloes with zero SFR, since in such calculations the SFR figures as a denominator, so our final dataset contains $112886$ subhaloes (Figure 1). 

\begin{figure}
\includegraphics[width=0.48\columnwidth]{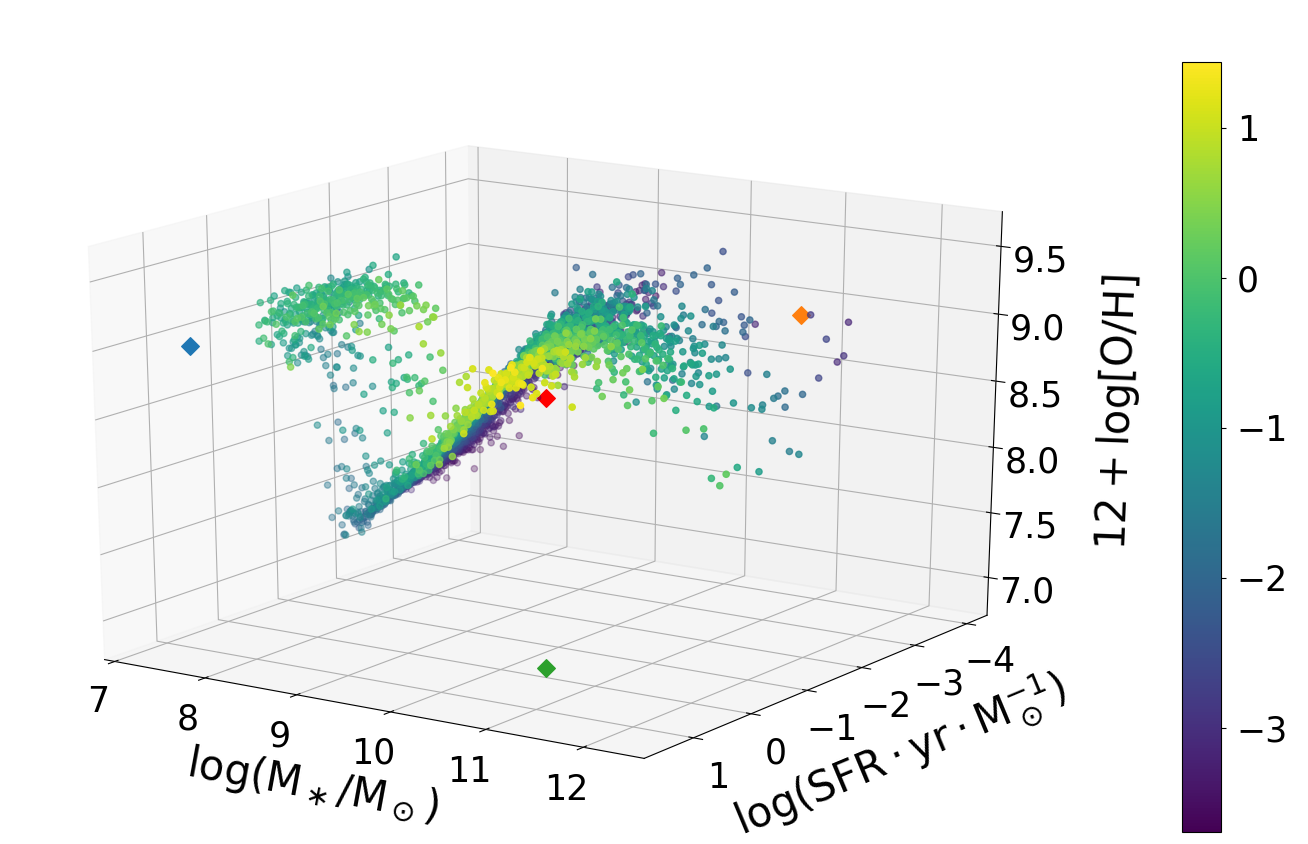}
\includegraphics[width=0.48\columnwidth]{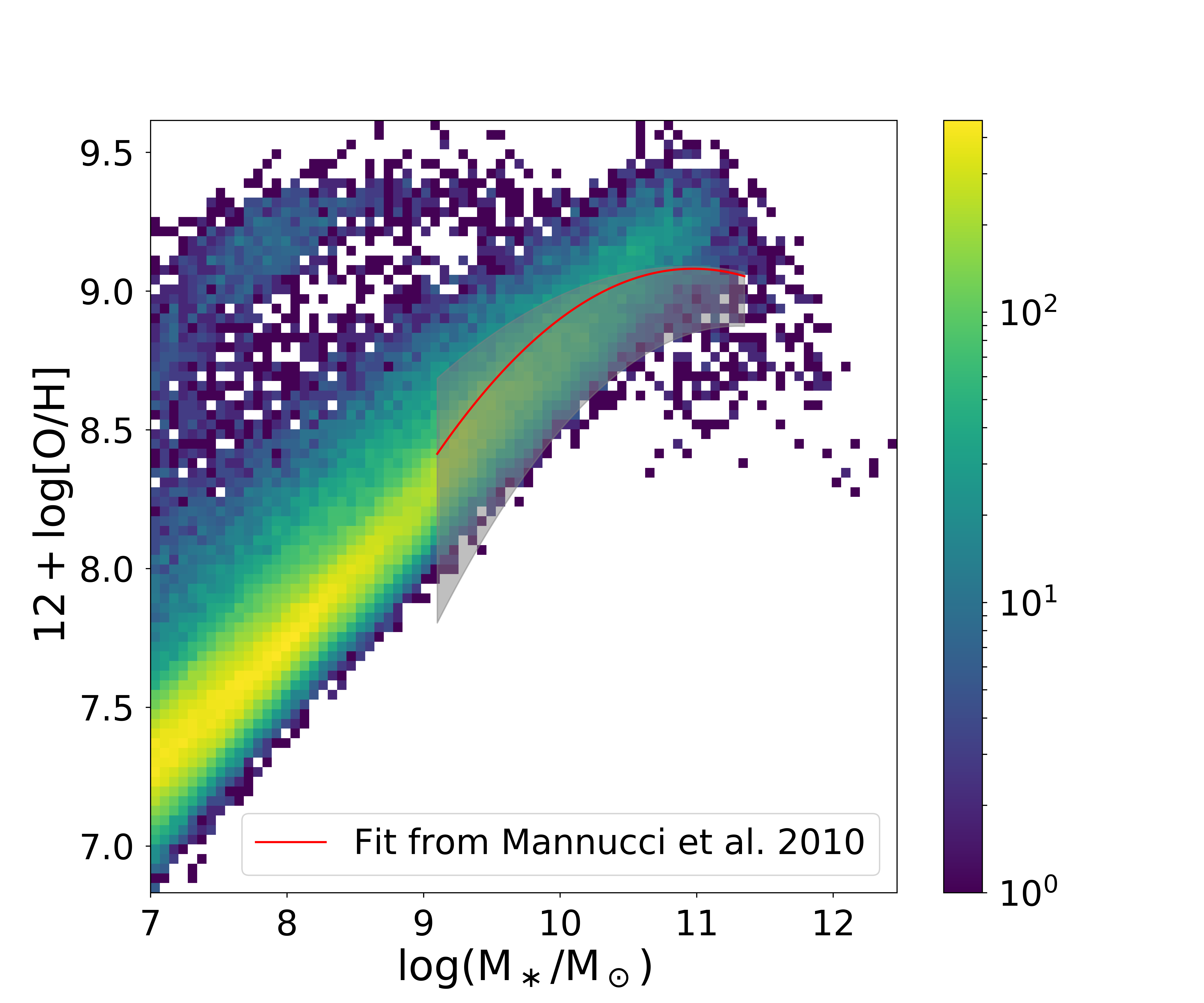}//
	\includegraphics[width=\columnwidth]{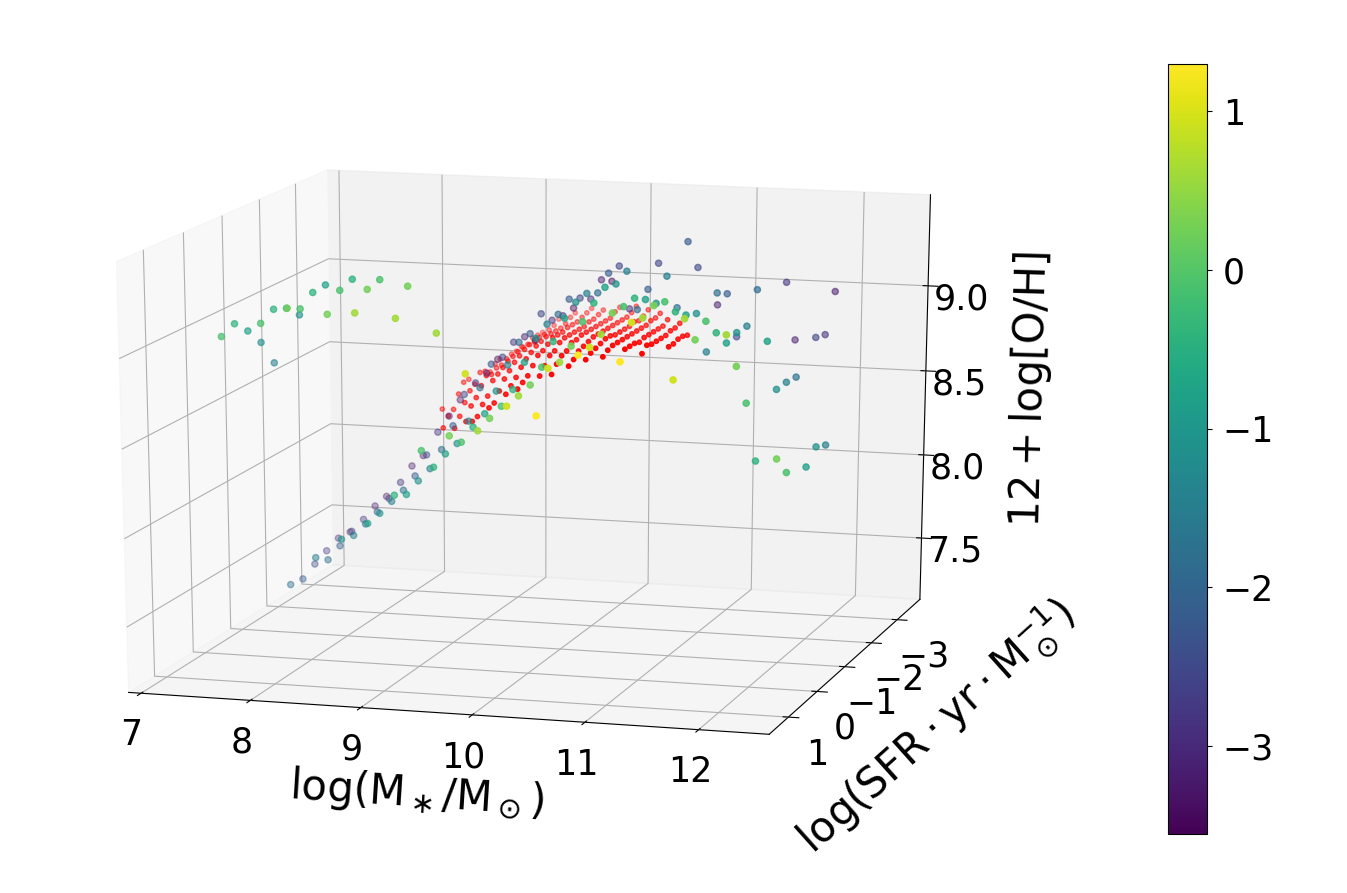}
    \caption{The total stellar mass, SFR, and gas-phase metallicity (metallicity of interstellar medium in a galaxy) for the selected sample of $112886$ subhaloes from the Illustris project. The Illustris metallicity data, given in units of mass fraction, is scaled using the constants from \citet{2017MNRAS.469.4921B}, to oxygen abundance in order to be comparable with the data from \citet{2010MNRAS.408.2115M}. Upper left:  The colour bar shows the values from the SFR axis. The Milky Way \citep[the values taken from][]{2015ApJ...810L...2D} is indicated with the orange diamond, and green, red and purple diamonds are its projections on the coordinate planes.  The data points are median values of metallicty in corresponding bins in the mass-SFR axis plane. Upper right: Colour-coded object count of the data projection along the SFR axis.  The red line shows the fit of Fundamental Metallicity Plane of galaxies from \citet{2010MNRAS.408.2115M}. The standard deviation of this plot is indicated with a shaded region around the curve.  Bottom:  Similar as the upper left panel. The median values are calculated with larger bin size so the data can be more clearly compared against the red dots \citep[the data from][]{2010MNRAS.408.2115M}.}
\end{figure}

Similar as in \citet{2010MNRAS.408.2115M} there is an obvious appearance of the Fundamental Metallicity Plane of galaxies.  The data presented in Figure 1 also shows a good agreement with the SDSS data presented in \citet{2004ApJ...613..898T}. The ILLUSTRIS data shows higher scatter across the plotted range of total stellar mass. For $\log M_* < 10$, both samples show a similar slope.  In the mass range of larger galaxies, the ILLUSTRIS data does not appear to have a shallower slope than the SDSS data. 

Apart from the Fundamental Metallicity Plane, the ILLUSTRIS data shows a sizeable subpopulation of dwarfs characterized by high stellar metallicity.  They constitute a ``cloudlet'' located above the fundamental plane, characterized by relatively low total stellar mass and low current star-formation rate.  This Cloudlet (designation we shall use in further text) contains $1617$ subhaloes. The Cloudlet separation from the Fundamental Metallicity Plane is evident in Figure 2.  From Figure 1, the part of the Cloudlet with the highest density of objects is located at $\log M_* \sim 8$ and in excess of $9.0$ (upper-right panel of Fig. 1) on the metallicity axis. This is also shown on Figure 2 as the strongest signal (blue line). For all three plotted mass scales in Fig. 2 there appears to be a separation between the Cloudlet and the Fundamental Metallicity Plane population.

\begin{figure}
\includegraphics[width=\columnwidth]{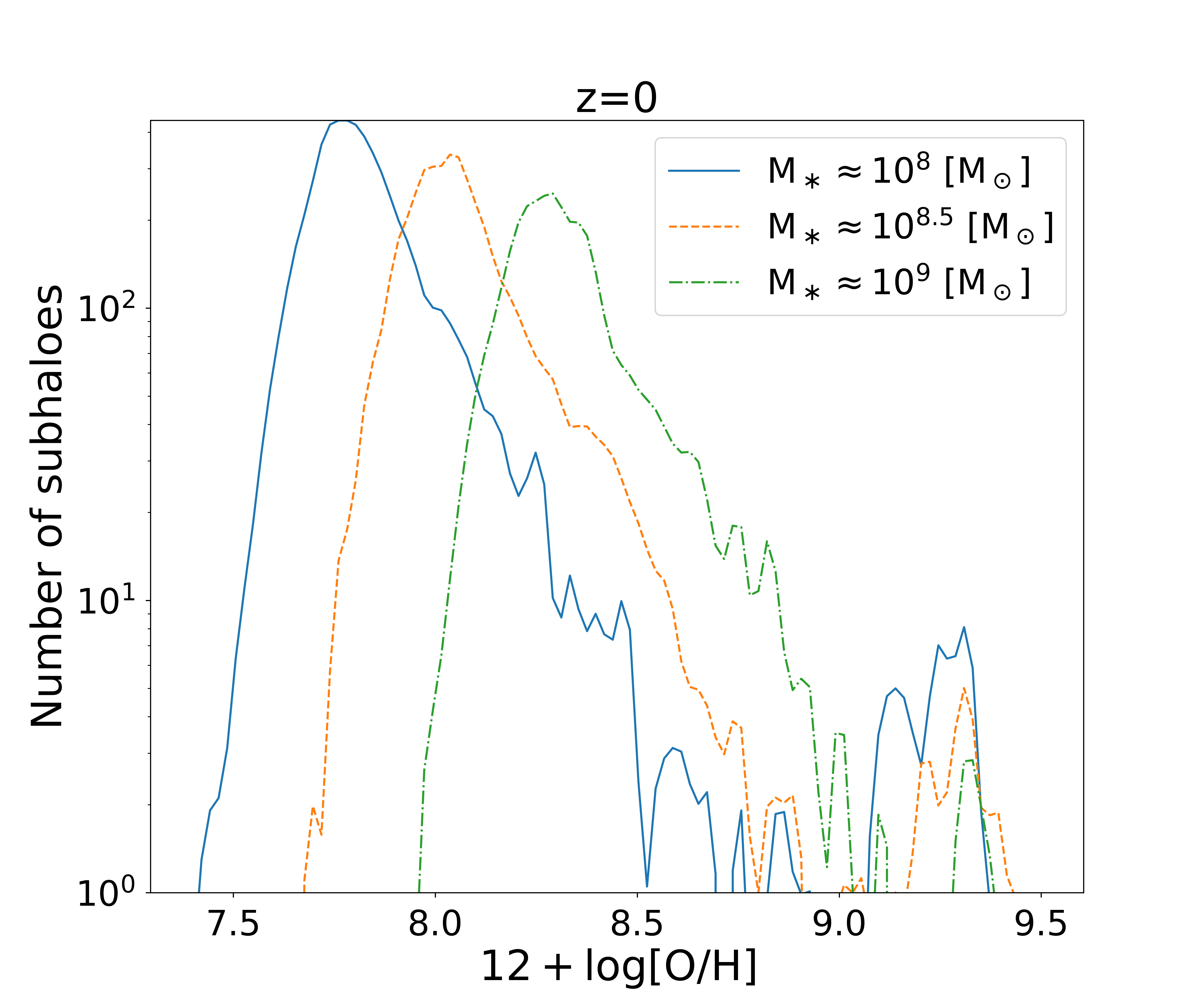}
    \caption{Number of subhaloes vs. oxygen abundance at specific values of halo mass (colour-coded as indicated). The  Cloudlet sample appears in form of separate peaks for the metallicity values above $9.0$. For all plotted masses, there is a  separation between the main sample and the Cloudlet indicative peaks.}
\end{figure}

\citeauthor{2014MNRAS.444.1518V} (\citeyear{2014MNRAS.444.1518V}, p. 1537ff) state that the physics used in the Illustris project simulation is explicitly complete and consistent only for objects with masses in excess of $10^9$ Solar masses but the majority of Cloudlet objects is below this limit (with only a handfull of more massive outliers, Figure 1,). However, it is important to keep in mind two things: (i) the limitation on the stellar mass claimed by the Illustris Project is a rather conservative one (there is still $\sim 10^5$ stellar particles even in the smallest resolved subhaloes); (ii) the explicit purpose in the limitation was studying morphological types of galaxies and comparison with the baryonic Tully-Fisher relationship. In fact, reliably identifying the type of a galaxy requires substantially more particles, and we will not attempt to quantify the morphological type of galaxies for systems that are resolved with less than about $10^5$ stellar particles \citep[][p. 1541]{2014MNRAS.444.1518V} In contrast, our study does not prejudicate morphological type of the Cloudlet objects -- we do not need all internal physics to be consistent. Essentially, especially for the late epochs close to $z$ = 0, we need just the metallicity-luminosity relationship to hold.  In the remainder of this paper we analyse the properties of objects belonging to this separate population in order to make the habitability assessment. It is not important that the detailed internal physics of such systems be consistent, only that the metallicity-luminosity relationship holds in the entire regime of interest \citep[cf.][]{2013ApJ...779..102K}.  

\begin{figure}
\includegraphics[width=\columnwidth]{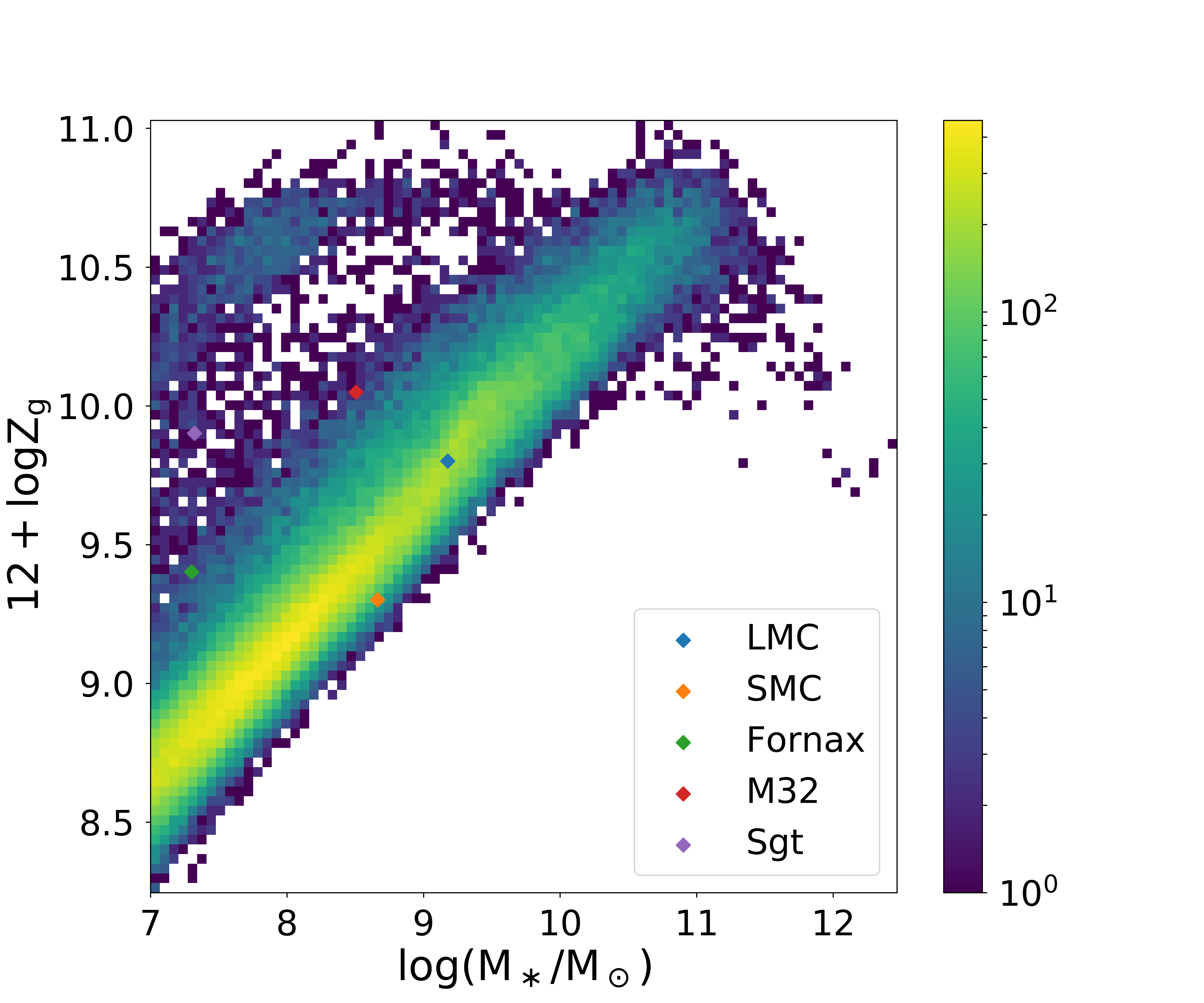}
    \caption{Colour-coded number of haloes for the data from the left panel of Fig. 1 in the metallicity -- mass plane. Some distinguished Local Group dwarfs \citep{2012AJ....144....4M} are plotted as indicated. }
\end{figure}

While some of the selected dwarf galaxies are in the Fundamental Plane, Fornax dwarf, and M32 appear to belong to the outskirts region of the Plane, towards the Cloudlet (see also Fig. 3). (The case of M32 may be more complex, if some recent views of its origin in tidal stripping are correct; we shall return to this topic in the concluding section.) The Sagittarius dwarf likely belongs to the Cloudlet population and with its mass $< 10^8 \,\mathrm{M}_\odot$ implies that existence of such a population is physically realistic even at the lower mass-end. In addition, the study of \citet{2014MNRAS.444.3684V} indicates -- although using a different kind of simulations -- that the metallicity-luminosity relationship for dwarfs continues to hold for objects with stellar masses $< 10^8\, \mathrm{M}_\odot$ and agrees well with the available observations (even for different types of dark matter particles). In addition, the similar slope of the ILLUSTRIS data Fundamental Metallicity Plane to the Mannuci et al. (2010) SDSS data, in the $9 < \log M_{*} < 10$ range, can be an indicator that the simulation data for Fundamental Metallicity Plane objects with masses below $\log M_{*}$ = 9 is in agreement  with the observations, despite the smaller consistency of implemented  physics at this mass range. This points to the overall acceptable fidelity of ILLUSTRIS data for metallicity, mass and star formation rate, for dwarf galaxies, including the Cloudlet objects.

\begin{figure}
\includegraphics[width=\columnwidth]{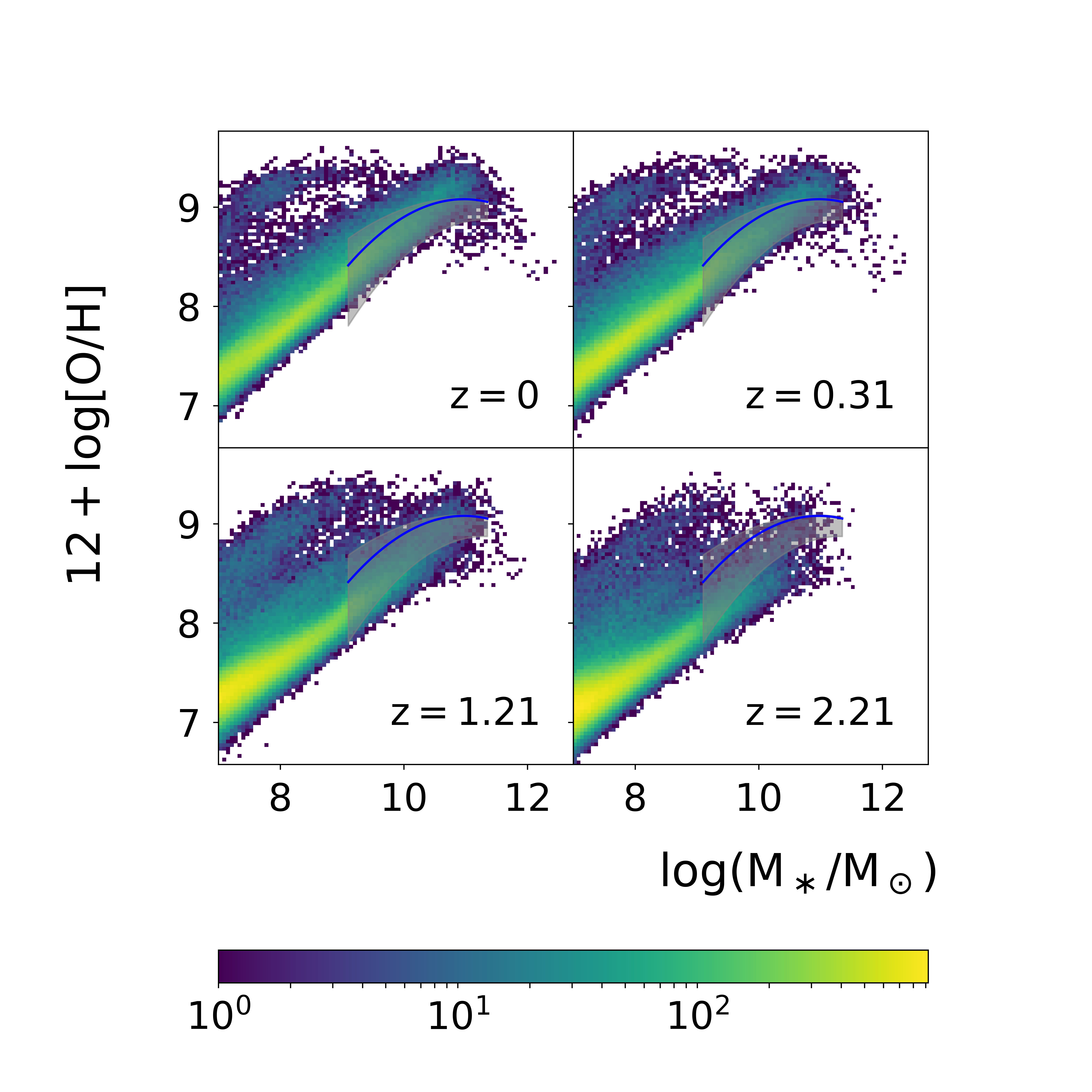}
    \caption{The Illustris data plot similar as in Figure 1 at different redshifts. The data from \citet{2010MNRAS.408.2115M} is also plotted as a blue curve with the shaded region indicating the uncertainty. The haloes count is colour-coded as indicated. }
\end{figure}

For further insight into the origin of the Cloudlet population, Figure 4 presents the plot similar to the one in Figure 1, but for different redshift epochs. At $z$ = 2.21 the Cloudlet is not present as a separate population, but rather as large dispersion of the main sample population. At $z$ = 1.21 the number of objects placed at the position of the Cloudlet seems to reach a maximum number and there is a small indication of the start of the population separation.  At $z$ = 0.3 there is a clear separation, with somewhat higher number of Cloudlet objects when compared to the present epoch. However, most of the Cloudlet objects are at the limit of ILLUSTRIS mass resolution during the simulated time span and consequently their merger history is not recorded. This is an important limitation of the ILLUSTRIS data because the objects that form the Cloudlet cannot be traced during the simulation, and the consistency of the Cloudlet history analysis is thus limited. The simulations with higher mass resolution are required to get more information about the Cloudlet sequence and shed more light on the physical reasons that led to the formation of the Cloudlet structure.

The highest density of objects in Figure 4 is present at higher redshifts, accompanied by the increased width of this density peak, compared to later epochs. However, for smaller redshifts the distribution is elongated towards higher mass end. The cosmic star formation rate has a peak value at $z \sim 2$. This is apparent at the Illustris data analysed in \citet{2014MNRAS.445..175G}, with the empirical confirmation in \citet{2013ApJ...770...57B}. The separation of the Cloudlet appears to start at $z < 2$, i. e., after the cosmic star formation rate starts to decline, which might present a part of the explanation for the high metallicity of the Cloudlet objects. 

The other part of the reasoning for high metallicity and low star formation rate might be related to the individual evolution of galaxies and their groups. In the review by \citet{2009ARA&A..47..371T},   two apparent evolution channels of dwarf galaxies are discussed; in one channel, galaxies lose their gaseous content significantly before than the ones in the other channel. In a recent work by \citet{2018MNRAS.479.1514B}, the authors distinguish between fast and slow dwarfs, with the fast ones exhibiting much more rapid star formation that results in subsequent loss of gaseous content caused by the supernovae gas heating feedback and possibly low dark matter content of these objects. The work also suggests the possible empirical insight into this evolutionary dichotomy with observations of shapes of the residual cores for these types of objects. Also, \citet{2015MNRAS.450.2367R} study young tidal dwarf galaxies, that have no dark matter, born in the tidal tails produced by interacting gas-rich galaxies. The authors distinguish between two populations of such objects; older population (likely originating at $z \sim 0.4$) that is formed from gas with lower metallicity and the recently-formed population that inherited higher metallicity from interacting larger galaxies. They compared their simulation data and found a good agreement with observations from \citet{2006ApJ...647..970L} for the former group, and with the observations from \citet{2010AJ....140.2124B} and \citet{2014MNRAS.440.1458D} for the latter. The objects of older population with masses $\sim 10^8\, \mathrm{M}_\odot$ have metallicity 12 + $\log[\mathrm{O/H}] \sim 8$ which corresponds to the upper part of the Fundamental Metallicity Plane data from our ILLUSTRIS sample (Figure 4). The more recently formed objects from the second group, with about $0.5$ higher value of metallicity, appear to fit into the transitional region between the Cloudlet and the main sample (Figure 4). This implies the possibility that the low metallicity end of the Cloudlet population might in part originate from this recently-formed tidal dwarf galaxies.  On the other hand, the two channels of dwarf galaxy formation discussed in \citet{2017MNRAS.470.4015M} imply that the Milky Way sized progenitors can be significantly striped to a dwarf sized objects when falling into the parent galactic cluster. This is consistent with the recent hypothesis of \citet{2018NatAs...2..737D} for the origin of M32, which will be discussed in Section 4 below.

These ideas imply that the separation between the Cloudlet and the main sample might be caused in part by the galactic mergers. \citet{2014MNRAS.445..175G} also examined the number density profiles at the cluster of galaxies. While the shape of the profiles does not change significantly with redshift it is apparent that the total number density decreases with time since galaxies exhibit mergers when sinking towards the centres of the clusters. Such circumstances might explain the reduction in the number of Cloudlet objects at the present epoch (Figure 4).

Further studies in this direction are essential, both theoretical and empirical. Figure 5 shows that Cloudlet objects are of the similar brightness as prominent dwarf galaxies.  Illustris simulation provides stellar photometrics for all of the simulated galaxies in eight bands (U, B ,V ,K, g, r, i, z) based on sum of the luminosities of the star particles within every subhalo. Using provided absolute magnitudes in V band, we calculated apparent magnitude of the cloudlet objects for three different distances (without the influence of the extinction): $10$, $100$ and $1000$ kpc. We used Pogson's formula, with absolute magnitude in V band taken from Illustris. Figure 5 illustrates that Cloudlet objects might well be observable like the other dwarf galaxies. However, the possible isolated nature of these objects might bring increasing difficulties as their search radius moves outside the Local group.

\begin{figure}
\includegraphics[width=\columnwidth]{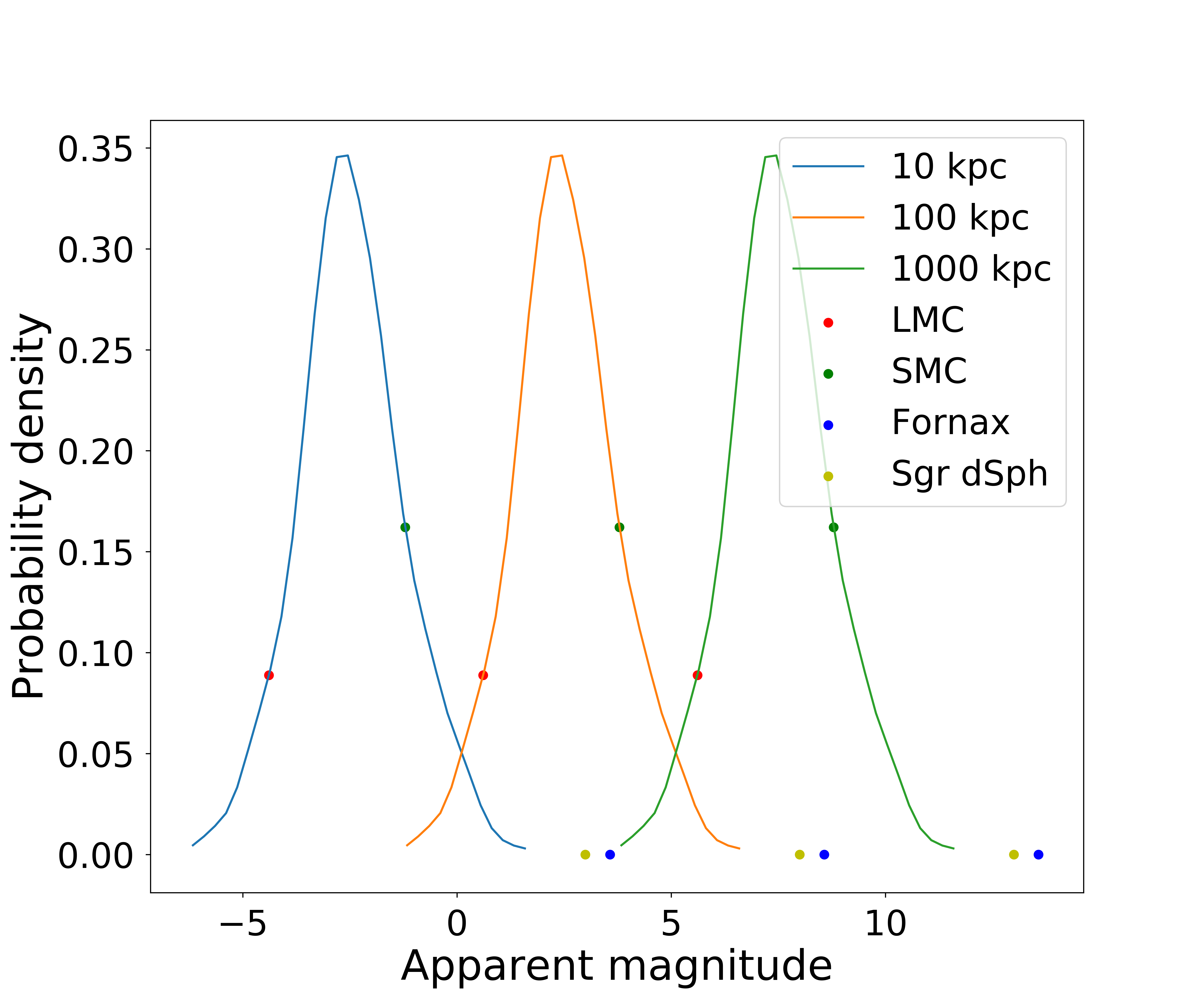}
    \caption{Apparent magnitude for the Cloudlet sample at different postulated distances. Fiducial Local Group dwarfs are shown as points for comparison. }
\end{figure}

\section{High-metallicity dwarfs as abodes of life}

The scatter in the metallicity-(stellar) mass relationship for the Cloudlet sample is very high, so it is difficult to obtain a regular power law, as discussed in detail by \citet{2013ApJ...779..102K}. The slope of such a power-law appears to be shallow. Our log-scale linear fit $y(x)$ = $ax$ + $b$ gives $a$ = $0.254 \pm 0.007$, $b$ = $8.420 \pm 0.054$. In comparison to the well-defined Fundamental Plane, Figures 1, 4, and 6 show scatter in the Cloudlet sample as significantly higher. The larger amount of scatter at the lower half of the considered mass interval might indicate that the cause of scatter can in part be attributed to the inconsistency of the simulation physics at these masses. 

\begin{figure}
\includegraphics[width=0.48\columnwidth]{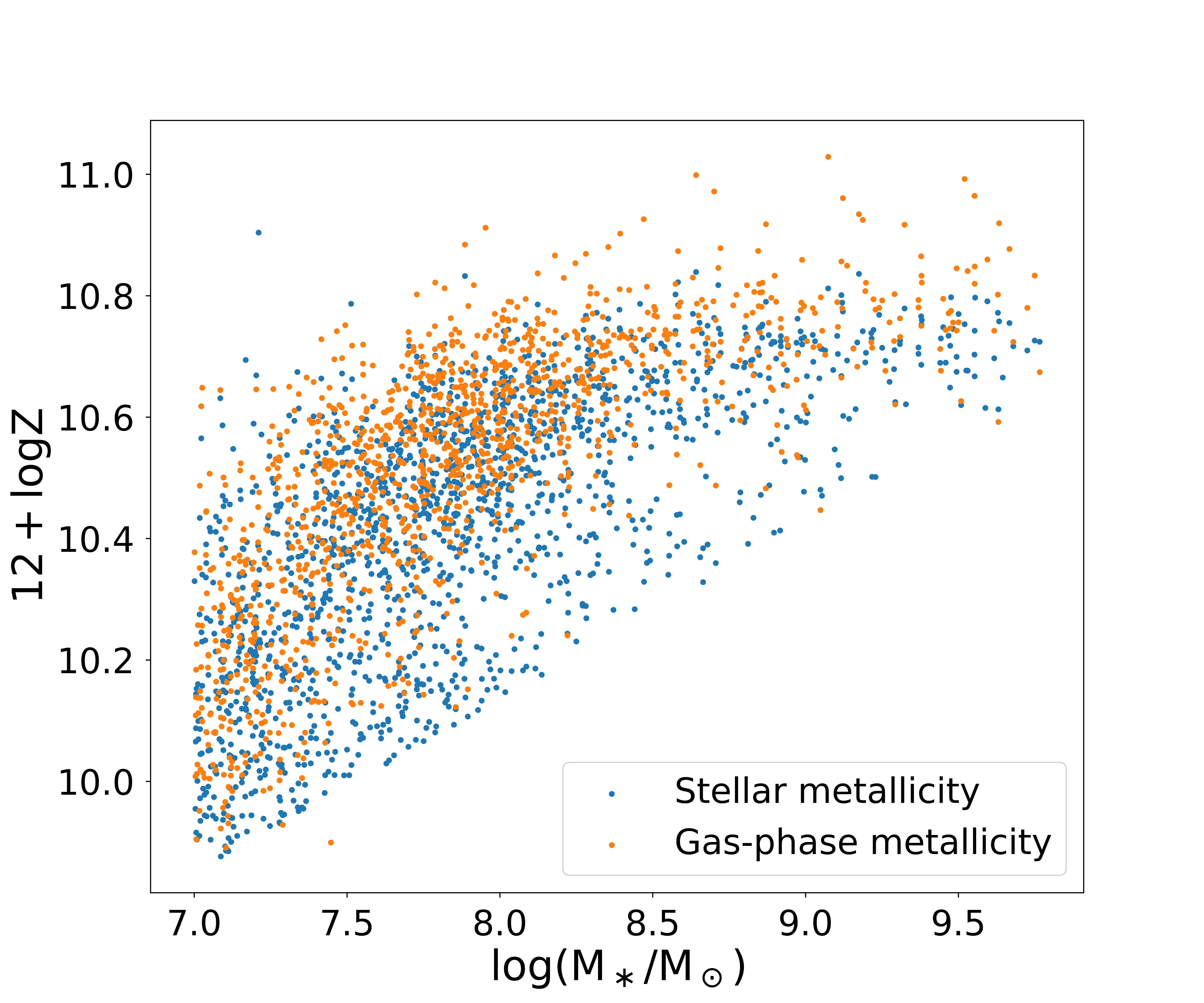}
\includegraphics[width=0.48\columnwidth]{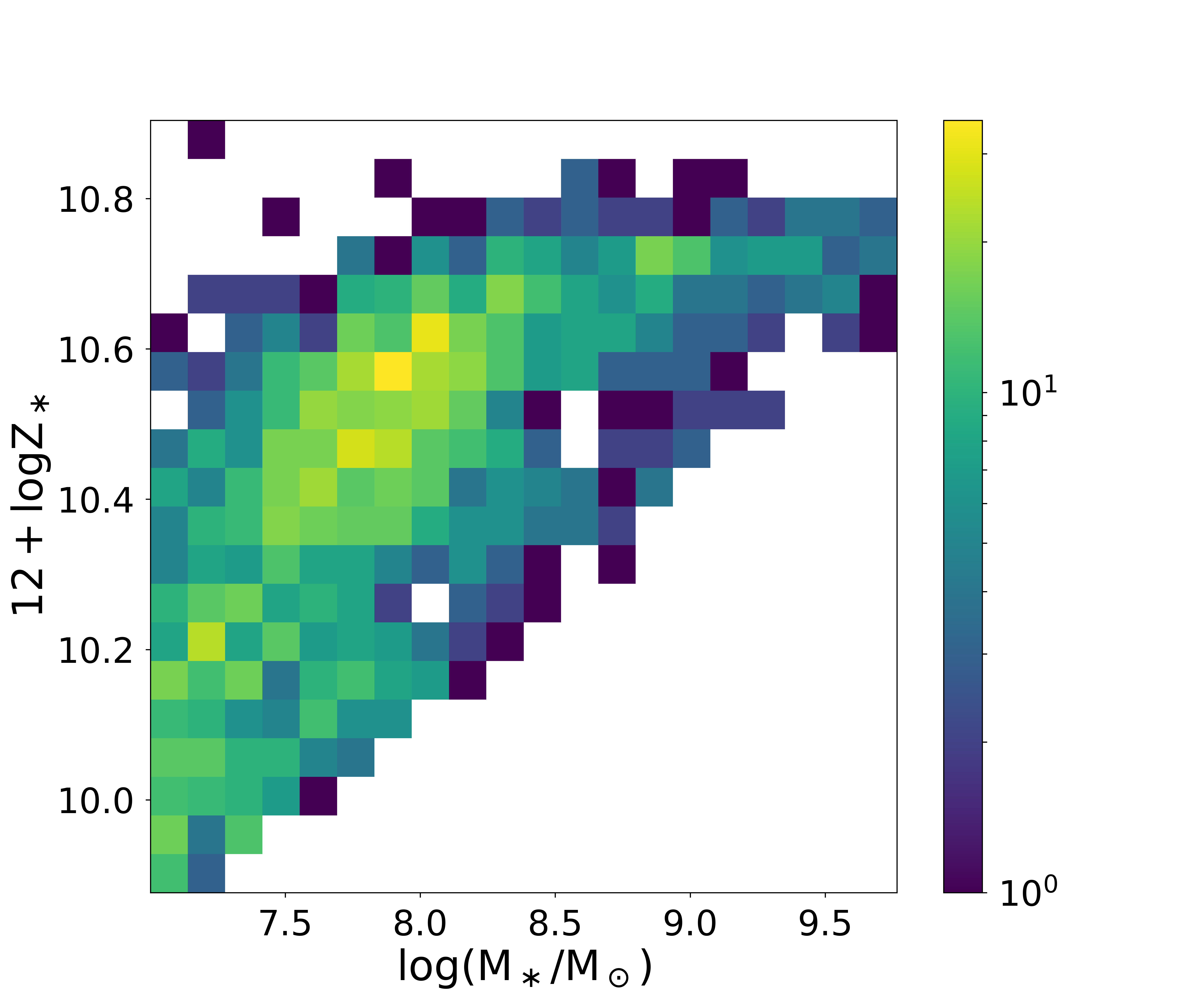}
    \caption{The Cloudlet of $1617$ subhaloes in the metallicity--mass plane. The left panel shows the correspondence between stellar and gas content while on the right is the colour-coded objects' count of the stellar component.  }
\end{figure}

Despite the uncertainties some provisional inference can be made. Figures 1 and 6 show preliminary indication of a metallicity--mass correlation in the dwarf regime which is similar to the Fundamental Plane. Indeed, the slope of the fit values from \citeauthor{2015ApJ...810L...2D} (\citeyear{2015ApJ...810L...2D}, esp. their Figure 3) is almost within the uncertainties of the Cloudlet slope obtained in the present study. In general, higher mass dwarfs tend to have higher metallicity and vice versa. The comparison presented in Figure 6 shows that this is likely the case for both stellar and gaseous components.

\begin{figure}
\includegraphics[width=0.48\columnwidth]{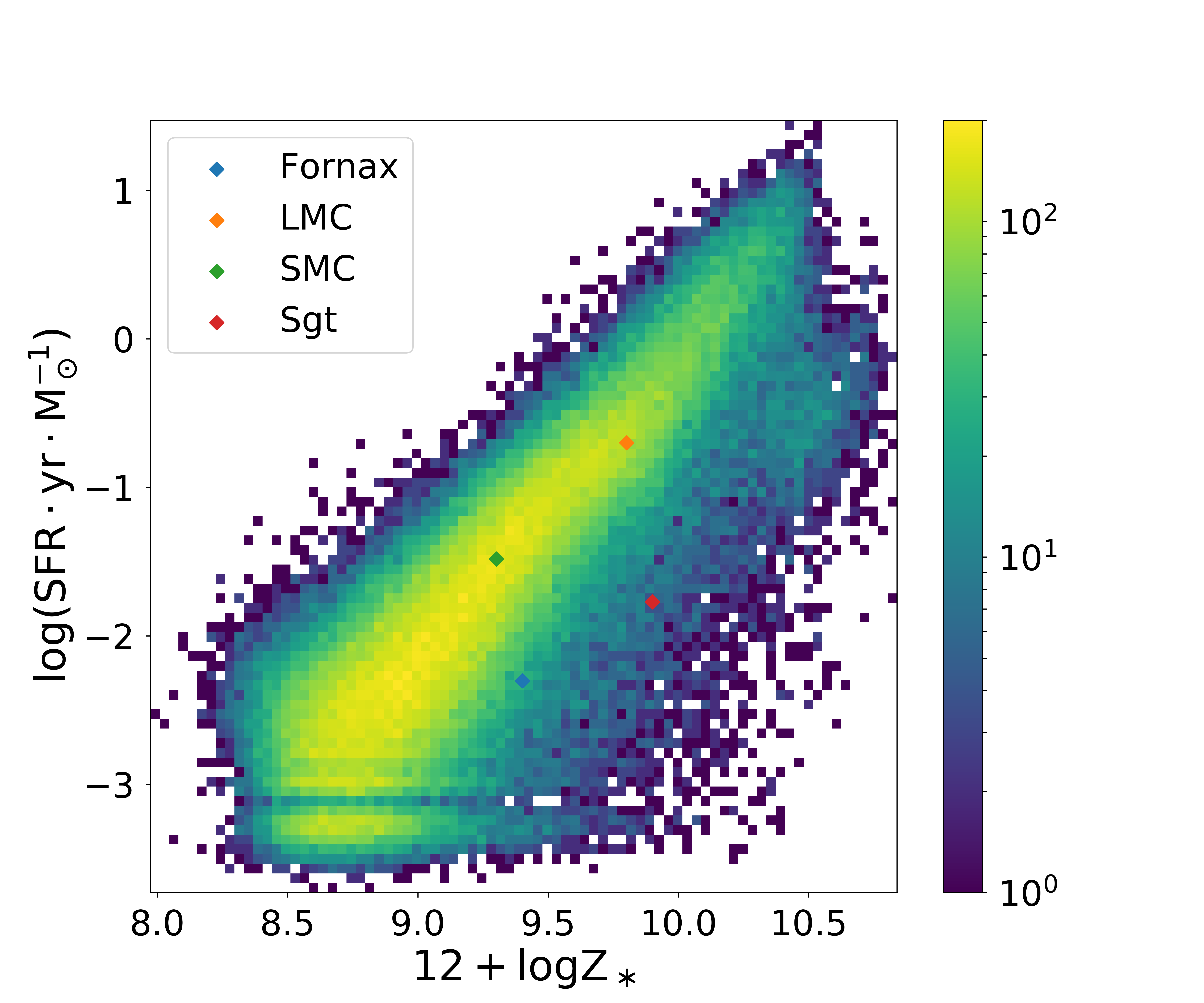}
\includegraphics[width=0.48\columnwidth]{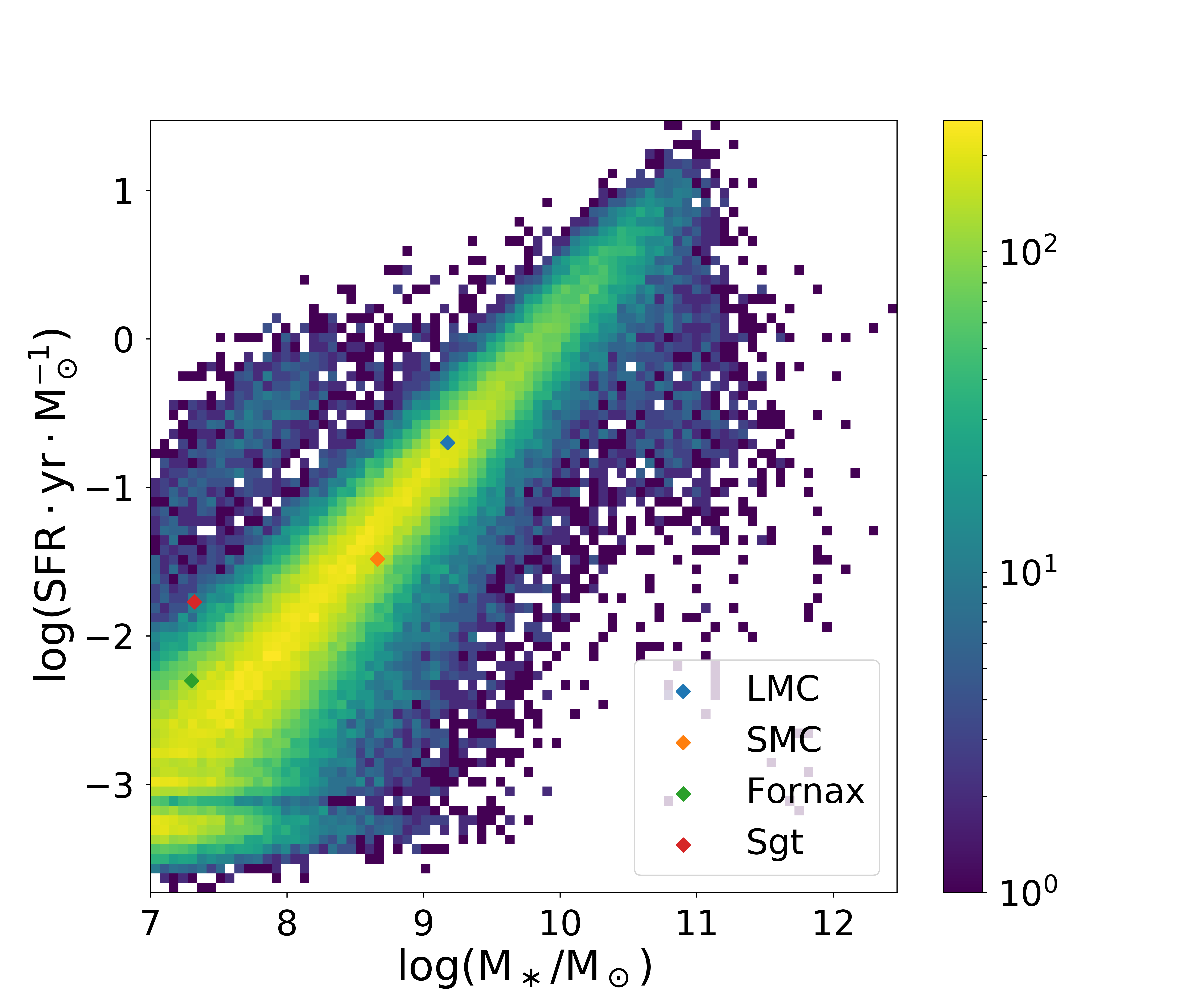}
    \caption{The colour-coded object count for the subhaloes sample from Figure 1 in the SFR--stellar mass plane (left) and SFR--metallicity plane (right). Prototypical dwarf galaxies from the Local group are indicated as in Figure 3. }
\end{figure}

\begin{figure}
\includegraphics[width=\columnwidth]{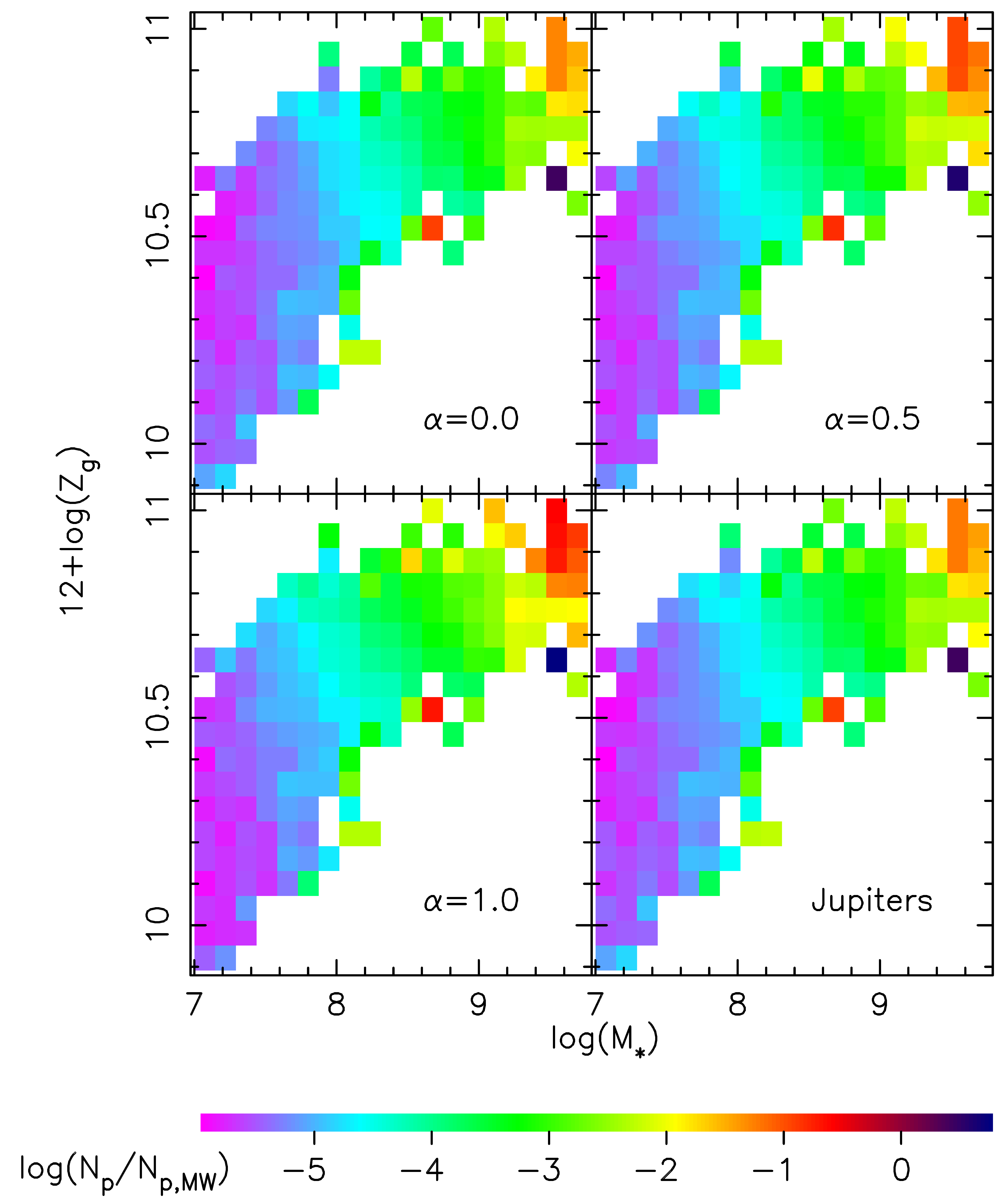}

    \caption{The habitability map for the Cloudlet objects. The number of habitable sites $N_\mathrm{p}$, normalized to the estimated number of planets for the Milky Way is presented with the colour bar. The colour-coded number of planets represents median values for the haloes from the Cloudlet sample. The number of terrestrial planets for each halo is calculated using Equation 5 from \citet{2015ApJ...810L...2D}, for different dependence on metallicity ($Z_\mathrm{g}^\alpha$). The number of Jupiter-like planets \citep[Eq. 4, ][]{2015ApJ...810L...2D} scales with metallicity as $10^{Z_\mathrm{g}}$.  }
\end{figure}

The significance of existence of such a correlation for the present research, as well as for the whole field of galaxy habitability, is further emphasized in Figure 7. While there is a general correlation for all objects comprising the Fundamental Plane, there is no separate correlation for the Cloudlet members; in fact, in both SFR--mass and SFR--metallicity diagrams we see correlations of distinctly lower quality as we go to smaller values of SFR (high scatter in the lower parts of both panels, especially in the SFR--metallicity relationship on the right-hand side). Any possible specific correlation for the population of high metallicity dwarfs should be taken with caution until both empirical scrutiny and simulation physics are improved. This in particular applies to observational constraints on this population, since it is very difficult to observe such intrinsically faint objects outside of the Local Group, which further entails the possibility of observation-selection effects biasing our conclusions; we shall return to this important topic in the concluding section. 

The quantity
\begin{equation}
N_\mathrm{p}\textrm{=}\begin{cases}
    \frac{M_{*}^2Z^\alpha}{\psi}, & \mathrm{terrestrial~planets}\\
    &\\
    \frac{M_*^210^Z}{\psi}, & \mathrm{moons~of~gaseous~giants}
  \end{cases}
\end{equation}
($\psi$ denoting SFR, $Z$ being the gaseous metallicity, and $\alpha$ an adjustable parameter expressing the scaling of probability for terrestrial planet formation with metallicity) serves as a proxy for the number of habitable terrestrial planets/moons of gaseous giants in \citet{2015ApJ...810L...2D}.  Figure 8 shows the habitability estimates as a number of the terrestrial planets relative to the Milky Way.  For the Milky Way we used mass of $6 \times 10^{10}\, \mathrm{M}_\odot$ and SFR of $3 \,\mathrm{M}_\odot/\mathrm{yr}$ \citep{2015ApJ...810L...2D} while the gaseous metallicity of $0.0196$ is taken from \citet{2017ApJ...839...55V}. 

The number of terrestrial habitable planets does not seem to have a strong dependence on metallicity \citep[Eq. 5 in ][]{2015ApJ...810L...2D}, at most approximately an order of magnitude for the given range of the parameter alpha. This is especially evident for the high mass and high metallicity objects. The number of Jupiter-like planets appears to be smaller even from the $\alpha$ = 0 case for the terrestrial planets. 

For a complete habitability assessment, it is also important to consider the continuous habitability time, as one of the basic requirements for the development and evolution of life. The limitations of the data used in this study (see Section 2 regarding the Figure 4), does not enable to trace the variables entering the equation for the number of habitable planets, such as the star formation rate or metallicity, as well as other factors (i.e.,  supernova rate) that might influence the habitability of individual Cloudlet objects throughout the simulation time span. While also important for understanding of the Cloudlet formation (Section 2), it is of crucial importance that further habitability studies have capability to identify lower mass individual objects and trace their evolution.  

It is obvious from Figure 8 that mass plays a dominant role for the number of planets per halo, similar as from the SDSS empirical assesment of \citet{2015ApJ...810L...2D}.  The pixels at the edge of the Cloud that have significantly higher number of  planets, than the corresponding inner part of the Cloudlet, have smaller SFR and consequently larger number of habitable planets. 

The absolute number of planets are nominally much smaller for a single dwarf galaxy, but since there are more dwarfs than giants, one should convolve these values with the corresponding distribution function. The baryonic version of the Schechter function may serve sufficiently well for stellar masses \citep{2003ApJ...585L.117B} in the local universe. One should be careful to distinguish the question ``which kind of galaxy hosts maximal number of habitable planets?'' from ``in which kind of galaxies do most of habitable planets reside?''

\section{Discussion: Two modes of habitability?}

Studies of galactic habitability are obviously in their infancy. There is a large number of galactic properties which may influence the habitability score in ways currently ill-understood, and even those whose influence is somewhat understood (like the mean metallicity or the global star-formation rate in the present study) are still only roughly represented in the models. While there is a huge space for further research, it is not too early to consider even rough hypotheses concerning galactic habitability, useful for guiding future work. Thus, we hereby propose an emerging picture of galactic habitability which is bimodal: high-metallicity dwarfs on one hand, and quiescent spiral discs on the other hand, represent the peaks of galactic habitability in the local universe at present.

The average metallicity of the Cloudlet members is $\sim 0.5$ dex higher than the metallicity of our Galaxy. Such a high metallicity value should enable more building material for planet formation but the dependence of Earth-like planets frequency on the metallicity of their host star is still debatable, largely due to a selection effect of detecting a larger Jupiter-like planets around higher metallicity stars. The influential work of \citet{2001Icar..151..307L} argued that although the probability of producing a terrestrial planet is growing with metallicity there is a sharp cut-off value at $> 0.3$ dex because larger metallicity leads also to production of larger, Jupiter-like planets (or oceanic ``super Earths'') that migrate towards the host stars and destroy smaller planets in the star's habitable zone. \citeauthor{2001Icar..151..307L} states that the $0.135$ dex is a peak of probability for hosting a habitable small planet, with a $68\%$ confidence range between -0.38 and +0.21 dex. Although a much smaller empirical sample of exo-planets was available at a time, the recent analysis of much larger samples did not significantly alter this picture. While the positive correlation between the host star metallicity and the frequency of Jupiter-like planets seems well established \citep{2018AJ....156..221N,2015AJ....149...14W,2012Natur.486..375B} the Earth-like planets occurrence rate is still debatable. While \citet{2012Natur.486..375B} did not report a significant dependence between the occurrence rate of the Earth-like planets and host star metallicity, \citet{2015AJ....149...14W} pointed that gas-dwarf and terrestrial-like planets are up to 2 times more frequent around stars with metallicity $> 0$ than around hosts with sub Solar metallicity values. The recent study of \citet{2019ApJ...873....8Z} argues that an average number of planets per star reaches a plateau at $> 0.1$ dex. The rationale behind this is that while Jupiter-like planets do appear more frequently at higher metallicities, they also seem to reside further away from the host star, as in our Solar System; this is obviously compatible with the existence of smaller rocky planets closer in.  In addition,\citet{2018ApJ...867L...3B} note that compact planetary systems appear as a constant fraction of planet-hosting stars even at high metallicity. Taken together with the present low SFR, and the corresponding very low supernova rates, the implication is that the Cloudlet members might be viable targets for astrobiological searches for biosignatures or SETI-related technosignatures.

However, further studies in this direction are needed, especially concerning the dynamical history of their stellar populations; as per cosmological GHZ simulations of \citet{2016MNRAS.459.3512V} and \citet{2017IJAsB..16...60F} processes such as radial diffusion, vertical oscillations, spiral-arm crossings, etc., are some of the key factors in the galactic evolution of habitability in spiral discs. The importance of similar processes, if they exist, in systems lacking the spiral disc-like component is unclear at present. The study of \citet{2008ApJ...685..904P} explores a sample of high-metallicity dwarfs including IC 225, finding them ``surprisingly nonpathological''. These objects, very similar to the Cloudlet members, seem morphologically undisturbed, gas-poor, and approaching the end of their star formation activity at present epoch. Their quiescence, coupled with high metallicity, clearly points to high relative degree of habitability. On the other hand, the dwarf galaxies are often companions to more massive counterparts and as such may have a rich, collision driven, dynamical history. Future simulations with improved physics at the low-mass end, coupled with the merger-tree analysis should illuminate some of these issues.

Insofar as there is a continuity between properties of low star formation rate/low metallicity stellar environments, a related topic is habitability of globular clusters in giant galaxies. Globular clusters have been the subject of astrobiological debates for quite some time \citep[especially if one counts the famous 1974 Arecibo message directed toward M13; cf. ][]{1996butc.book.....D}. As in the case of early-type galaxies, the ``first wave'' (pre-Kepler) astrobiology discarded any habitability in globular clusters due to metallicity concerns \citep[][and subsequent GHZ studies until recently]{2001Icar..152..185G,2005OLEB...35..555G}. \citet{2012A&A...546L...1D} argue that persistence of habitable planets is strongly suppressed in the globular cluster environment due to frequent planetary ejections in close stellar encounters. In contrast, \citet{2016ApJ...827...54D} find that there are stable ``sweet spots'' within globular clusters where terrestrial habitable planets could form and provide suitable habitable environments. While obviously more work needs to be done to ascertain the habitability status of the globular cluster environment, it seems that things are not clear-cut here either, allowing for at least some terrestrial, transiently habitable planets, further eroding the ``rare Earth'' dogma that only high-metallicity, Pop I stellar environments support habitability. 

If the emerging picture is correct, one mode of habitability is based upon the following factors: large galaxies with active star formation, similar to the Milky Way, with habitable planets forming around the Pop I analog stars with high metallicities. This mode is characterized by many different kinds of planetary systems, a high level of chemical evolution and, presumably, easier routes for advancement of prebiotic chemistry in both interstellar/circumstellar medium and on planetary surfaces. Once life appears, however, it is subject to strong abiotic selection pressure of its astrophysical environment in form of frequent irradiations by supernovae and GRBs, perturbations due to the spiral-arm and galactic-plane crossings, higher cosmic-ray fluences, and other astrobiological regulation mechanisms \citep{2008SerAJ.176...71V}. Part of these perturbing influences may overflow into the ``Gaian windows'' \citep{2016AsBio..16....7C}, which are likely to be shorter for biospheres in the giant spiral systems. 

The other mode of habitability takes place within lower-star formation systems, characterized by lower (but not prohibitively low) metallicity and much smaller irradiation frequency and other perturbations. It is reasonable to assume that abiogenesis happens more seldom there, but once life emerges, it is more likely to persist, take root and spread. There are indications of possible fundamental lower limit (``floor'') of mean stellar metallicity in dwarf systems \citep{2012AJ....144....4M}. How big is the scatter of individual stellar (hence planetary) metallicities about those minimal values determines whether potentially habitable planets exist in the low-metallicity part of the distribution of galaxies. 

This emerging bimodal picture then suggests that high-metallicity dwarfs might, in addition to giant ellipticals like the Maffei 1, be good place for practical searches. This is in complete accordance with the results of the cosmological simulation of \citet{2017IJAsB..16...60F}. As far as nearby systems are concerned, the Large Magellanic Cloud, M32, Sagittarius dSph, Canis Major, M110/NGC 205, IC 225, and the Fornax dwarf represent targets of interest for future extragalactic biosignature and technosignature searches. Some of them have possibly very interesting history: compact ellipticals might emerge as a consequence of tidal stripping of giant spirals \citep[e.g., ][]{2017MNRAS.470.4015M} and this in particular might apply to M32 as has been recently suggested by \citet{2018NatAs...2..737D}. What are the implications of such evolutionary trajectory for habitability remain to be seen, since the hypothetical stripping in the case of M32 occurred 2 Gyr ago; if these events happen early enough, the low-irradiation regime is firmly established by the last snapshot at $z$ = 0. Further research, focused on evolutionary histories and environmental impact on dwarf galaxies, is necessary to determine how frequent are such ``interlopers'' and how big a weight they should be assigned in the overall average habitability calculation.  

It is interesting to speculate that hypothetical advanced technological civilizations arising in dwarf galaxies would have an easier route to dominating their stellar systems than those arising in giants like the Milky Way. Hence, new and innovative research programs such as the G-hat survey for high Kardashev Type civilizations \citep{2014ApJ...792...26W,2014ApJ...792...27W,2015ApJS..217...25G} or the search for dynamical-mass outliers \citep{1999JBIS...52...33A,2015ApJ...810...23Z} should be extended to these kinds of objects; obviously, the latter method would need to be modified to use correlates of the dynamical mass other than the Tully-Fisher relation.

A hypothetical further reason why dwarf galaxies may be the best abodes of life consists in shortness or absence of the AGN phase as far as their central engines are concerned. Since an early, visionary hypothesis of \citet{1981Icar...46...94C}, there has been some concern whether the irradiation by the nuclear source could play a significant role in galactic habitability; \citeauthor{1981Icar...46...94C} actually suggested that LMC and SMC might be more habitable than the Milky Way on this basis. A recent study by \citet{2017NatSR...716626B} indicates that the Milky Way nuclear source could adversely influence hypothetical biospheres in a few central kiloparsecs. In particular, atmospheric loss due to XUV irradiation during the AGN phase of the Milky Way is total within 1 kpc, and some adverse consequences for complex life could be expected up to 10 kpc. In contrast, \citet{2018ApJ...855L...1C} obtain much smaller radii of influence of the central source, on the order of only $20-30$ pc, although their study is focused on the atmospheric evaporation of Neptune-like objects, which could leave Earth-like solid cores. These studies have not yet been generalized to external galaxies. In particular, the scaling of nuclear activity in terms of intensity and duration at small values of M$_*$ should be studied by approach similar to that of \citet{2016MNRAS.461.3322M}. While isolated dwarf galaxies might be an obvious habitability ``sweet spots'', interacting dwarf galaxies could also have a significant habitability potential. A prolonged period of AGN in dwarf galaxy might be the result of slow long-lasting gas accretion triggered by gravitational interaction with surrounding galactic structures. This scenario supports our second mode of habitability, since according to \citet{2019MNRAS.487.5549B}, the AGN feedback might quench the star formation rate rendering the galaxy more habitable at later epochs. This is an important topic for further work.

Finally, there is a problem of observation-selection effects. Per analogy with the reasoning of \citet{2018IJAsB..17...77H} in connection to stars having habitable planets, we might wish to ask why we find ourselves in a giant spiral galaxy instead of an early-type galaxy and/or a dwarf. In contrast to the stellar case, we do not know the final states of galaxy evolution, so it is not appropriate to weight likelihood by temporal duration \citep{cirkovic_balbi}. One strategy is, obviously, to reject the anthropic reasoning and accept that the Milky Way is an exceptional fluke, in particular by being too quiet \citep{2007ApJ...662..322H} in terms of merger- and star-formation histories \citep[cf. ][]{2009A&A...505..497Y}. The discovery of Fermi bubbles \citep{2010ApJ...724.1044S} casts doubt on this conclusion, however. Even if the conclusions of \citet{2007ApJ...662..322H} are correct, the subset of $(7 \pm 1)\%$ spiral galaxies found to be similar to the Galaxy in a sample of well-studied local systems is not really that minuscule a fraction that anthropic reasoning should trouble us much; it is incomparable with other known fine-tunings where the observed values fall within a fraction of the parameter space equal to $10^{-5}$ or less \citep{2000RvMP...72.1149H}. Other approaches are possible, however; it is possible that further research will indicate that there is a temporal cut-off to the habitability of galaxies, perhaps through insofar neglected dynamical aspects of stellar migrations within a galaxy of particular type \citep[see the discussion in ][]{2016MNRAS.459.3512V}. Alternatively, it is possible that our astrobiological view of the local universe is skewed too much by particular properties of the Local Group, including both the Milky Way and the specific dwarfs mentioned above. So far, there are no observational indications to the extent that the Local Group is indeed typical in a larger set of similar small galaxy groups at $z$ = 0. Finally, we may entertain the idea that it is not habitability in general, but only the reference class of observers like us, which has an ending in a finite future epoch and therefore has a finite and relatively small measure. These are just some of many open questions which motivate further research at the confluence of astrobiology, cosmology, and futures studies.

\section*{Acknowledgements}

We thank an anonymous referee for many helpful comments that have significantly improved the manuscript. The authors acknowledge financial support from the Ministry of Education, Science and Technological Development of the Republic of Serbia through the project \#ON176021 ``Visible and invisible matter in nearby galaxies: theory and observations''.

\bibliography{stojkovic}
\end{document}